\documentclass[%
reprint,
superscriptaddress,
nofootinbib,
amsmath,amssymb,
aps,
prc,
floatfix, ]
{revtex4-2}

\usepackage{graphicx}
\usepackage{dcolumn}
\usepackage{bm}
\usepackage[bookmarks=true,colorlinks,%
            citecolor=blue,linkcolor=blue,anchorcolor=blue,filecolor=blue,urlcolor=blue,%
           ]{hyperref}
\allowdisplaybreaks

\newcommand{\bra}[1]{\langle {#1} |}
\newcommand{\ket}[1]{| {#1} \rangle}

\newcommand{\vecr}{\mathbf{r}}
\newcommand{\vecR}{\mathbf{R}}
\newcommand{\psihat}{\hat{\psi}}

\begin{document}


\title{
Local \texorpdfstring{$\alpha$}{alpha}-removal strength in the mean-field approximation
}

\author{Takashi Nakatsukasa}%
 \affiliation{Center for Computational Sciences,
              University of Tsukuba, Tsukuba 305-8577, Japan}
 \affiliation{Faculty of Pure and Applied Sciences,
              University of Tsukuba, Tsukuba 305-8571, Japan}
 \affiliation{RIKEN Nishina Center, Wako 351-0198, Japan}
\author{Nobuo Hinohara}%
 \affiliation{Center for Computational Sciences,
              University of Tsukuba, Tsukuba 305-8577, Japan}
 \affiliation{Faculty of Pure and Applied Sciences,
              University of Tsukuba, Tsukuba 305-8571, Japan}
 \affiliation{Facility for Rare Isotope Beams, 
              Michigan State University, East Lansing, Michigan 48824, USA}

\date{\today}

\begin{abstract}
\noindent{\bf Background:}
The $\alpha$ cluster is a prominent feature,
not only in light nuclei but also in heavy nuclei.
To study the $\alpha$-particle formation
in the mean-field calculation,
the localization function has been
extensively utilized.
However, the localization function does not guarantee
the proximity of four different nucleons
which is required by the $\alpha$-particle formation.
A simple indicator of the proximity is desired.
Recently, experimental measurement of the quasifree $\alpha$-knockout reaction for Sn isotopes
reveals the cross sections with a monotonic decrease 
with increasing neutron number.
[Science {\bf 371}, 260 (2021)].
This is interpreted as evidence of
the surface $\alpha$ formation.
\\
\noindent{\bf Purpose:}
We propose a simple and comprehensible quantity to assess the proximity of
four nucleons with different spins and isospins.
Using this, we examine the recent measurement of
$\alpha$-knockout cross sections in Sn isotopes.
\\
\noindent{\bf Methods:}
The local $\alpha$-removal strength is proposed
to quantify the possibility to form an $\alpha$ particle
at a specific location inside the nucleus.
In addition, it provides the strength of ground and excited states
in the residual nuclei after the removal of the $\alpha$ particle.
To make the calculation feasible,
we introduce several approximations,
such as point-$\alpha$, mean-field, and no rearrangement approximations.
We use the Hartree-Fock-plus-BCS method for
the mean-field calculation for Sn isotopes.
We also propose another measure, the local $\alpha$ probability,
which should provide a better correlation with the $\alpha$-knockout cross sections.
\\
\noindent{\bf Results:}
The calculation of the local $\alpha$-removal strength
is extremely easy in the mean-field model with no rearrangement.
For even-even Sn isotopes,
the local $\alpha$-removal strengths to the ground state of residual nuclei
are almost universal in the nuclear surface region.
In contrast,
the local $\alpha$ probability
produces strong neutron number dependence consistent with
the experiment.
\\
\noindent{\bf Conclusions:}
The local $\alpha$-removal strength
and the local $\alpha$ probability are easily calculable
in the mean-field models.
Recent experimental data for Sn isotopes
may be explained
by a simple model without explicit consideration of $\alpha$ correlation.
\end{abstract}


\maketitle


\section{Introduction}

Clustering is an intriguing phenomenon in the nuclear structure.
Correlations between nucleons result in the formation of
subunits (clusters) inside the nucleus.
The most typical cluster is the $\alpha$ particle,
which is present not only in light nuclei,
but also observed in heavy nuclei as the $\alpha$-decay phenomena.
In light nuclei, prominent clustering often takes place in
excited states whose energy is close to the threshold of
the corresponding cluster decomposition \cite{Ikeda_diagram68}.

The microscopic theories of the clustering phenomena have a long history
\cite{Whe37-1,Whe37-2,Ikeda_diagram68,PTPS_cluster_review80,FHKLM18}.
In fact, Gamow's theory of the $\alpha$ decay \cite{Gam28,Gam30} was
published even before the discovery of the neutron \cite{Cha32}.
Most theoretical studies of the cluster structure in the past have been
performed with an assumption that a certain cluster structure exists
in the nucleus.
It is common to construct the cluster wave functions
in terms of the Gaussian wave packets \cite{Brink66}.
For instance,
the antisymmetric molecular dynamics (AMD) \cite{Hor92}
and the fermionic molecular dynamics (FMD) \cite{Fel89} were
extensively utilized in studies of the nuclear cluster phenomena
and heavy-ion reactions.
In the AMD and FMD,
the cluster structure is not assumed {\it a priori},
although the Gaussian wave packet is assumed for a single-particle state
\cite{KKO12,FS97,FS00}.
The configuration mixing, which is often treated
with the projection and the generator coordinate method \cite{HW53,GW57},
plays an important role in the studies of clustering in relatively light nuclei.
In the AMD and FMD,
since each Gaussian has parameters corresponding to the position of its center
and the magnitude of its width,
the clustering can be identified by
close location of centers of many Gaussian wave packets.

In contrast, the mean-field (energy density functional) theory
can provide the optimal single-particle wave functions
to minimize the total energy of a Slater determinant.
One of the advantageous features of
the theory is the capability of describing almost all the nuclei
in the nuclear chart using a single energy density functional
which is a functional of normal and pair densities.
Another advantage can be the treatment of the pairing correlations,
which become indispensable
especially for heavy nuclei with open-shell configurations.
The obtained (generalized) single-particle states are, in most cases,
spread over the entire nucleus, not confined in a localized region
inside the nucleus.
Therefore, in the mean-field theory, it is not straightforward
to find nucleons’ gathering in terms of the single-particle wave functions.
In relatively light nuclei, prominent cluster structure
could be observed by
the nucleon density profile \cite{TYI96,OYN04,MKSRHT06,MLIK10,Fuk13}.
In particular, the clustering phenomena have been extensively studied
with relativistic energy density functionals \cite{EKNV17}.
The cluster structure in the relativistic energy density functionals,
such as DD-ME2, is more visible
than in the nonrelativistic functionals,
producing spatially localized subunits inside the nucleus \cite{EKNV12}.
However, the identification of the cluster structure has some ambiguity
and relies on one's intuition.

There exist some methods aiming to identify and quantify the clustering effect
in the mean-field states.
In case one can intuitively build a model cluster wave function,
its overlap with the mean-field state gives
a possible measure of the clustering \cite{MKSRHT06}.
Recently, another method,
which does not require a model wave function,
has been proposed to visualize
possible cluster correlations using the mean-field wave functions
\cite{MT22}.
However, the application of the method seemingly become more and more difficult 
as the nucleon number increases.

The localization function, introduced into nuclear physics
by Reinhard and collaborators \cite{RMUO11},
is a possible measure of $\alpha$-particle formation.
Similar functions were introduced in molecular physics
to investigate the shell structure and the chemical bonding \cite{BE98}.
Since it only needs one-body densities,
such as kinetic and current densities,
the calculation requires negligible computational cost.
In addition, it is given as a function of the spatial coordinates.
Thus, one can identify the location of the $\alpha$ particles.
Because of these advantageous properties,
the localization function has been adopted in a number of studies
for the cluster correlations within the mean-field theory 
\cite{ZSN16,SN17,IM18,Tan19,MBNEKV21,RVNZZM22}.
However, it should be noted that
the localization function does not examine whether the four nucleons
exist next to each other.
It tells us information on
the conditional pair density for particles of the same kind,
$P_{q\sigma}(\vecr,\vecr')$
where $q=n,p$ and $\sigma=\pm 1/2$.
Reference~\cite{RMUO11} clearly states that
the localization function is just the first step
to identifying the $\alpha$ cluster.
The $\alpha$ cluster requires the four nucleons
to gather in a localized region inside the nucleus,
which cannot be checked by the localization function.
Properties of the localization function were studied
in Ref.~\cite{KHME22},
concluding that it is not sensitive to the compactness of the $\alpha$ particle.
Therefore,
the purpose of the present paper is
to propose the next step, ``local $\alpha$-removal strength,''
as a measure of {\it four localized nucleons}
that can be easily estimated in the mean-field theory.

Experimentally, the $\alpha$ correlations in nuclei can be 
investigated by quasifree $\alpha$-knockout reactions \cite{CRCNC81,RCCGHW77}.
A recent experiment on the $\alpha$-knockout reactions in Sn isotopes
by Tanaka and collaborators \cite{Tan21}
reveals that the cross section monotonically decreases as
the neutron number increases.
They interpret this trend as a tight interplay between
the $\alpha$ formation and the neutron skin \cite{Typ14}.
The distorted-wave impulse-approximation study shows that
the reaction takes place in a peripheral region
and probes the $\alpha$ particles in the nuclear surface
\cite{YMO16}.
Another purpose of the present paper is to examine
consistency between the calculated local $\alpha$-removal strength
and the result of Ref.~\cite{Tan21}.

The paper is organized as follows:
We propose a feasible measure of four-particle localization,
local $\alpha$-removal strength, in Sec.~\ref{sec:local_alpha_strength_function}.
In Sec.~\ref{sec:results},
the local $\alpha$-removal strength is applied to Sn and other isotopes.
The calculation is compared with the measurement of the $\alpha$-knockout reaction.
Concluding remarks are given in Sec.~\ref{sec:conclusion}.

\section{\label{sec:local_alpha_strength_function} Local \texorpdfstring{$\alpha$}{alpha}-removal strength}


\subsection{Definition}

Let us assume a single Slater-determinant description for
the $\alpha$ particle and that
the orbital part of the single-particle wave functions
are all the same and given by $\phi_\alpha(\vecr)$,
where the center of mass of the $\alpha$ particle is located at the origin.
Then, the $\alpha$-particle annihilation operator $\hat{\alpha}(\vecR)$
at the position $\vecR$ is given by
\begin{align}
	\hat{\alpha}(\vecR)&\equiv
	\int
 	d\vecr_1
	d\vecr_2
	d\vecr_3
	d\vecr_4
        \phi_\alpha^*(\vecr_{1\vecR})
	\phi_\alpha^*(\vecr_{2\vecR})
	\phi_\alpha^*(\vecr_{3\vecR})
	\phi_\alpha^*(\vecr_{4\vecR})
	\nonumber \\
	& \quad \times\psihat_\uparrow^{(n)}(\vecr_1)
	\psihat_\downarrow^{(n)}(\vecr_2)
	\psihat_\uparrow^{(p)}(\vecr_3)
	\psihat_\downarrow^{(p)}(\vecr_4)
	\label{eq:alphahat_1}
\end{align}
where $\vecr_{i\vecR}=\vecr_i-\vecR$ ($i=1,\ldots,4$)
and $\psihat^{(q)}_{\sigma}(\vecr)$
indicates the field operator for the particle of the isospin $q=n,p$ and
the spin $\sigma=\uparrow,\downarrow$.
The wave function $\phi_\alpha(\vecr)$ is a well-localized function,
normally assumed to be a Gaussian in the cluster model.

\begin{figure}[t]
	\centerline{
	\includegraphics[width=0.3\textwidth]{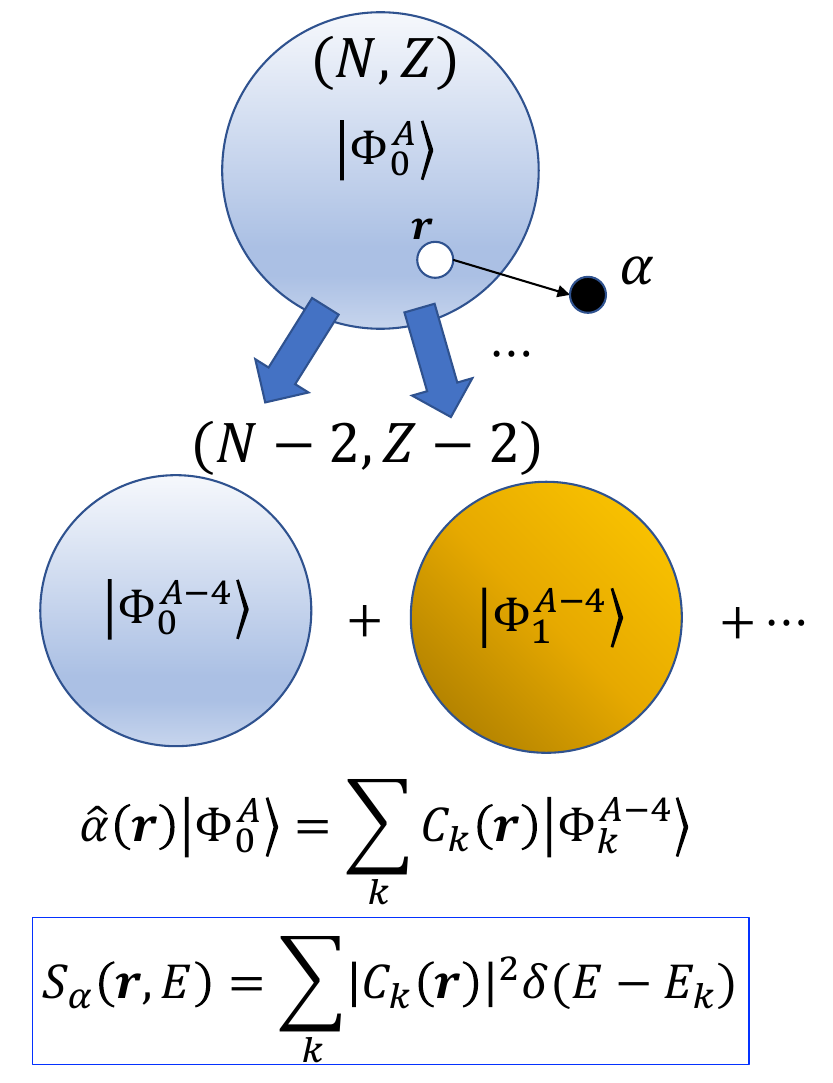}
}
	\caption{
		Schematic illustration of 
		the local $\alpha$-removal strength $S_\alpha(\vecr,E)$.
		Removing the $\alpha$ particle located at a position $\vecr$
		from the nucleus $(N,Z)$
		results in many energy eigenstates of the residual nucleus
		$(N-2,Z-2)$, with
		$C_k(\vecr)=\bra{\Phi^{A-4}_k} \hat{\alpha}(\vecr) \ket{\Phi_0^A}$.
		See Eq. (\ref{eq:S(r,E)}).
	}
	\label{fig:ponchi}
\end{figure}
To investigate the $\alpha$ particle in the nucleus,
we propose ``local $\alpha$-removal strength'' defined as
\begin{equation}
	S_\alpha(\vecr,E)\equiv \bra{\Phi_0^A} \hat{\alpha}^\dagger(\vecr)
	\delta(E-\hat{H}) \hat{\alpha}(\vecr) \ket{\Phi_0^A} ,
	\label{eq:definition_of_a}
\end{equation}
where $\hat{H}$ is the Hamiltonian,
and 
$\ket{\Phi_0^A}$ is the ground state of the nucleus $(N,Z)$.

The meaning of this quantity is clear if we insert
the unity expanded in terms of the complete set
for the nucleus $(N-2,Z-2)$, $\{ \ket{\Phi_k^{A-4}} \}$:
\begin{equation}
	S_\alpha(\vecr,E)=\sum_{k=0}^\infty
	\left|\bra{\Phi^{A-4}_k} \hat{\alpha}(\vecr) \ket{\Phi_0^A}\right|^2
	\delta(E-E_k^{A-4}) ,
	\label{eq:S(r,E)}
\end{equation}
where $\hat{H}\ket{\Phi_k^{A-4}}=E_k^{A-4}\ket{\Phi_k^{A-4}}$.
Thus, the quantity
\begin{eqnarray}
{\cal S}_\alpha(\vecr)_{E,\Delta E}
	&=& \int_{E-\Delta E/2}^{E+\Delta E/2} S_\alpha(\vecr,E') dE' \nonumber \\
	&=& \sum_{k}^{\Delta E}
	\left|\bra{\Phi^{A-4}_k} \hat{\alpha}(\vecr) \ket{\Phi_0^A}\right|^2
	.
\label{eq:S_EDeltaE}
\end{eqnarray}
provides the strength of the transition to states in the energy range
$(E-\Delta E/2,E+\Delta E/2)$
of the residual nucleus,
when the $\alpha$ particle is removed at the position $\vecr$
in the nucleus $(N,Z)$.
See also Fig.~\ref{fig:ponchi}.

It is convenient to define the variable $E$ with respect to
the ground-state energy $E_0^{A-4}$,
namely, the excitation energy $E' = E-E_0^{A-4}$.
Hereafter, we denote $E'$ as $E$ for simplicity.
Appropriate smearing of the $\delta$ function $\delta(E-E_k^{A-4})$
may be useful for visualizing the strength as a function of $E$.
The local $\alpha$-removal strength $S_\alpha(\vecr,E)$
may provide transition strength to states at the excitation energy $E$,
when the $\alpha$ particle is removed at the position $\vecr$.

\subsection{Approximations}

The calculation of Eq. (\ref{eq:definition_of_a})
demands a large computational cost in general.
We introduce here some approximations
to make the computation feasible.

\subsubsection{Point \texorpdfstring{$\alpha$}{alpha} approximation}

First, 
in order to avoid the multiple integrations in Eq. (\ref{eq:alphahat_1}),
we approximate the wave function $\phi_\alpha(\vecr)$
by the $\delta$ function $\delta(\vecr-\vecR)$.
Thus, in this paper, we use
\begin{equation}
	\hat{\alpha}(\vecr)=
	\psihat^{(n)}_{\uparrow}(\vecr)
	\psihat^{(n)}_{\downarrow}(\vecr)
	\psihat^{(p)}_{\uparrow}(\vecr)
	\psihat^{(p)}_{\downarrow}(\vecr) .
	\label{eq:alphahat}
\end{equation}

This approximation significantly reduces the computational cost. 
Without the point $\alpha$ approximation, the numerical cost of the
multiple integrals with respect to the four coordinates is
extremely large.
Since the Gaussian wave function for the $\alpha$ particle is
compact, the point $\alpha$ approximation is able to provide
a useful signal of the localized four nucleons.
Further approximations in the following sections
lead to products of pair densities
$\langle \psihat^{(q)}_\uparrow(\vecr_1)\psihat^{(q)}_\downarrow(\vecr_2)\rangle$.
Without the point $\alpha$ approximation,
we need nonlocal pair densities with $\vecr_1\neq\vecr_2$,
whose behavior is not well controlled
in currently available pairing energy density functionals.

\subsubsection{Mean-field approximation}

Next, we adopt the mean-field ground state for $\ket{\Phi_0^A}$,
and the Hamiltonian $\hat{H}$ is approximated in the mean-field level.
$\hat{H}$ is truncated up to the second order
in terms of the quasiparticle (qp) operators defined with respect
to the ground state of the residual nucleus $(N-2,Z-2)$:
\begin{equation}
	\hat{H}=\sum_{q=n,p}\sum_{i>0}
	E_i^{(q)} \hat{a}_i^{(q)\dagger} \hat{a}^{(q)}_i
	+ \cdots ,
	\label{eq:mean-field-Hamiltonian}
\end{equation}
where the ground-state energy of the nucleus $(N-2,Z-2)$,
$E_0^{A-4}=\bra{\Phi_0^{A-4}}\hat{H}\ket{\Phi_0^{A-4}}$,
is subtracted.
$\hat{a}^{(q)}_i$ and $E^{(q)}_i$ are the qp annihilation
operators and corresponding qp energies.
The subscript $i>0$ means the summation with respect to
the qp states with positive qp energies $E^{(q)}_i>0$.
In Eq. (\ref{eq:S(r,E)}),
the excited states ($k>0$) are
given by an even number of qp excitations.
Thus, the index $k$ stands for 2qp, 4qp, $\cdots$
and the excitation energies $E_k^{A-4}$ ($k>0$) in Eq. (\ref{eq:S(r,E)}) are given as
\begin{eqnarray}
E_{ij,0}^{A-4}&=&E^{(n)}_i+E^{(n)}_j ,
\label{eq:E_n2qp} \\
E_{0,ij}^{A-4}&=&E^{(p)}_i+E^{(p)}_j ,
\label{eq:E_p2qp} \\
E_{ij,i'j'}^{A-4}&=&E^{(n)}_i+E^{(n)}_j+E^{(p)}_{i'}+E^{(p)}_{j'} ,
\label{eq:E_4qp}
\end{eqnarray}
and so on.

With the mean-field construction of the states,
$\ket{\Phi_0^A}$
and $\ket{\Phi_0^{A-4}}$,
one can calculate the transition matrix elements
$\bra{\Phi^{A-4}_k} \hat{\alpha}(\vecr) \ket{\Phi_0^A}$
in Eq. (\ref{eq:S(r,E)})
as follows.
Except for the cases under the presence of
proton-neutron ($pn$) pairing \cite{BV63,Goo72,PRDN04}
and/or $pn$ mixing \cite{SDNS13,She14},
the states are normally described by product wave functions
of protons and neutrons,
$\ket{\Phi^A}=\ket{\Phi^N}\otimes\ket{\Phi^Z}$.
Therefore, the transition matrix elements can be also written in the
product form,
$\bra{\Phi^{N-2}_{k}} \psihat^{(n)}_\uparrow(\vecr)
\psihat^{(n)}_\downarrow(\vecr) \ket{\Phi_0^N}
\bra{\Phi^{Z-2}_{k'}} \psihat^{(p)}_\uparrow(\vecr)
\psihat^{(p)}_\downarrow(\vecr) \ket{\Phi_0^Z}
$,
where $k$ ($k'$) stands for 0qp, 2qp, $\cdots$ indices
for neutrons (protons).
Thus,
\begin{equation}
	S_\alpha(\vecr,E)
	=\sum_{k\geq 0}\sum_{k'\geq 0}
	F_{k}^{(n)}(\vecr) F_{k'}^{(p)}(\vecr)
	\delta(E-E_{kk'}^{A-4}) ,
	\label{eq:S(r,E)_MF}
\end{equation}
where $E_{kk'}^{A-4}$ are given by
Eqs. (\ref{eq:E_n2qp})$-$(\ref{eq:E_4qp}), 
and
\begin{equation}
	F_k^{(q)}(\vecr)=
\left| \bra{\Phi^{N_q-2}_k} \psihat^{(q)}_\uparrow(\vecr)
\psihat^{(q)}_\downarrow(\vecr) \ket{\Phi_0^{N_q}} \right|^2,
	\label{eq:F_k}
\end{equation}
with $N_q=N$ and $Z$ for $q=n$ and $p$, respectively.

\subsubsection{Neglect of rearrangement}

We introduce further approximation to neglect the rearrangement
of the mean fields due to the removal of the $\alpha$ particle.
Hence, we assume that the mean fields in nuclei of mass number
$A$ and $A-4$ (before and after the removal of an $\alpha$ particle)
are identical.
When the neutrons (protons) are in a superfluid phase,
we also neglect the change of the chemical potential,
which leads to $\ket{\Phi_k^{N_q-2}}\approx\ket{\Phi_k^{N_q}}$
with $q=n$ ($p$).
With this approximation,
the calculation of the residual states ($\ket{\Phi_k^{A-4}}$)
is no longer required.
The mean-field Hamiltonian (\ref{eq:mean-field-Hamiltonian})
is now replaced by that for the nucleus $(N,Z)$,
in which all the quasiparticle states are defined with respect
to the mean-field ground state of $\ket{\Phi_0^A}$.

Assuming the Bogoliubov transformation \cite{RS80},
\begin{eqnarray}
	\hat{a}_i^\dagger&=&\sum_\sigma\int d\vecr \left\{
	U_i(\vecr\sigma)\psihat_\sigma^\dagger(\vecr)
	+V_i(\vecr\sigma)\psihat_\sigma(\vecr) \right\}, \\
\psihat_\sigma^\dagger(\vecr)&=&\sum_{i>0} \left\{
	U_i^*(\vecr\sigma) \hat{a}_i^\dagger
	+ V_i(\vecr\sigma)\hat{a}_i \right\},
\end{eqnarray}
the matrix elements $F^{(q)}_k(\vecr)$ of Eq. (\ref{eq:F_k}) are given as
\begin{eqnarray}
	F_0(\vecr) &=& \left|\sum_{i>0} U_i(\vecr\uparrow)V_i^*(\vecr\downarrow)
	\right|^2
	=\left|\kappa(\vecr)\right|^2,
	\label{eq:F_0_super}\\
	F_{ij}(\vecr) &=& \left| V_i^*(\vecr\uparrow)V_j^*(\vecr\downarrow)
	- V_j^*(\vecr\uparrow)V_i^*(\vecr\downarrow) \right|^2,
	\label{eq:F_ij_super}
\end{eqnarray}
where the superscript $(q)$ is omitted for simplicity. 
$F_0(\vecr)$ is nothing but a square of the local pair density,
$|\kappa(\vecr)|^2\equiv|\bra{\Phi_0^A}\psihat_\uparrow(\vecr)
\psihat_\downarrow(\vecr)\ket{\Phi_0^A}|^2$.
It should be also noted that, with this approximation,
only the 0qp and 2qp excitations of neutrons and protons
contribute to the summation
with respect to $k$ and $k'$ in Eq.~(\ref{eq:S(r,E)_MF}).

For the transition to the ground state,
which may be of most interest,
the calculation is feasible with these approximations.
Its relative values among different isotopes
may be a useful indicator of the $\alpha$-particle knockout probability.
However, we should keep in mind that
this is based on the approximations adopted,
and should be careful especially when we compare the values
for nuclei in different mass regions.

\subsubsection{HF-plus-BCS approximation}

Using the HF-plus-BCS (HF+BCS) approximation,
the HFB wave functions are proportional to the HF single-particle states
$\{ \phi_i(\vecr\sigma) \}$ as
\begin{equation}
	\begin{split}
	&U_i(\vecr\sigma)=u_i \phi_i(\vecr\sigma), \quad
	V_i(\vecr\sigma)=-v_i \phi^*_{\bar{i}}(\vecr\sigma) , \\
	&U_{\bar{i}}(\vecr\sigma)=u_i \phi_{\bar{i}}(\vecr\sigma), \quad
	V_{\bar{i}}(\vecr\sigma)=v_i \phi^*_i(\vecr\sigma) ,
	\end{split}
\end{equation}
where $\phi_{\bar{i}}$ is the time-reversal partner of $\phi_i$.
The BCS uv factors, $(u_i,v_i)$, are all real and determined by
the HF single-particle energies \cite{RS80}.
This recasts Eqs.(\ref{eq:F_0_super}) and (\ref{eq:F_ij_super})
into
\begin{eqnarray}
	F_0(\vecr) &=& \left|\kappa(\vecr)\right|^2
	=\left|\sum_i u_i v_i 
		\phi_{\bar{i}}(\vecr\uparrow)\phi_i(\vecr\downarrow) \right|^2
	\label{eq:F_0_BCS}\\
	F_{ij}(\vecr) &=& v_i^2 v_j^2 \left| \phi_i(\vecr\uparrow)\phi_j(\vecr\downarrow)
	-\phi_j(\vecr\uparrow)\phi_i(\vecr\downarrow) \right|^2 .
	\label{eq:F_ij_BCS}
\end{eqnarray}
The summation in Eq.~(\ref{eq:F_0_BCS}) is taken over
both $i$ and $\bar{i}$
with $\phi_{\bar{\bar{i}}}=-\phi_i$, $u_{\bar{i}}=u_i$, and $v_{\bar{i}}=v_i$.\footnote{
Explicitly denoting the time-reversal parts,
Eq.~(\ref{eq:F_0_BCS}) can be written as
\begin{equation}
	F_0(\vecr) =
	\left|\sum_{i\gg 0} u_i v_i \left\{
		\phi_{\bar{i}}(\vecr\uparrow)\phi_i(\vecr\downarrow)
		-\phi_i(\vecr\uparrow)\phi_{\bar{i}}(\vecr\downarrow) \right\}
	\right|^2.
\end{equation}
Here, $i\gg 0$ indicates the summation is not taken
over $\bar{i}$.
Note that it is different from $i>0$ in Eq.~(\ref{eq:mean-field-Hamiltonian}).
}
Since the indices $ij$ and $ji$ correspond to the same 2qp excitation,
the summation in Eq.~(\ref{eq:S(r,E)_MF}) is performed with
respect to different combinations of 2qp indices,
namely with the restriction of $i>j$.

The pairing gap $\Delta_q$ ($q=n,p$) is related to
the monopole pairing strength $G$ as
$
\Delta_q=\frac{1}{2} G\sum_i u_i v_i
$
\cite{RS80}.
In this paper, we adopt the value of $\Delta_q$
as the experimental odd-even mass difference.

\subsubsection{No pairing case}

In case there is no pairing (normal phase)
in the state $\ket{\Phi_0^{N_q}}$, 
Eq.~(\ref{eq:F_0_super}) does not give a transition
to the ground state because the pair density trivially vanishes.
This is due to the wrong approximation of
$\ket{\Phi_k^{N_q-2}}\approx\ket{\Phi_k^{N_q}}$ for the normal phase.
In this case, 
$\ket{\Phi_0^{N_q}}$ is
the Hartree--Fock (HF) ground-state wave function.
Then, we explicitly remove two particles from the occupied orbitals in
$\ket{\Phi_0^{N_q}}$, and identify it as $\ket{\Phi_k^{N_q-2}}$
in Eq. (\ref{eq:F_k}).
This leads to
\begin{equation}
	F_{ij}(\vecr) = \left| \phi_i(\vecr\uparrow)\phi_j(\vecr\downarrow)
	- \phi_j(\vecr\uparrow)\phi_i(\vecr\downarrow) \right|^2,
	\label{eq:F_ij_normal}
\end{equation}
where $\phi_i$ and $\phi_j$ are the single-particle wave functions
for the occupied (hole) states.
The expression is equal to Eq. (\ref{eq:F_ij_super})
by identifying $V_i^*=\phi_i$ ($V_i^*=0$)
for hole (particle) states.
Note that $ij$ are the two-hole indices
which include not only the excited states ($k>0$) but also the ground state
($k=0$) in $\ket{\Phi_k^{N_q-2}}$.
For the ground state $\ket{\Phi_0^{N_q-2}}$, 
two particles $ij$ are removed from the highest occupied orbitals (HOO).

The ground state of the residual nucleus $(N-2,Z-2)$ is unique
when the ground state of the nucleus $(N,Z)$ is superfluid
both in protons and neutrons.
However,
we should remark here that,
for the normal state (no pairing),
there may be multiple ground states with the present approximations.
This is an undesired consequence of no rearrangement, and
it occurs when the nucleus $\ket{\Phi_0^A}$ is spherical,
because the HOO with the angular momentum $j$
should have a degeneracy of $2j+1$.
To keep the feasibility in the computation,
we simply sum up all the possible two-hole indices
to produce $F_0$ for nuclei in the normal phase.
\begin{equation}
	F_0(\vecr)=\sum_{ij\in {\rm HOO}} 
	\left| \phi_i(\vecr\uparrow)\phi_j(\vecr\downarrow)
	- \phi_j(\vecr\uparrow)\phi_i(\vecr\downarrow) \right|^2.
	\label{eq:F_0_normal}
\end{equation}

\subsection{Localization function}

We will compare the local $\alpha$-removal strength with
the localization function $C_\sigma^{(q)}(\vecr)$
in Sec.~\ref{sec:results}.
It may be useful to recapitulate the definition and the meaning of
$C_\sigma^{(q)}(\vecr)$, according to Ref.~\cite{RMUO11}.

The conditional probability of finding a nucleon
with spin $\sigma$ and isospin $q$ at $\vecr'$
when another nucleon with the same spin and isospin
exists at $\vecr$ is given by
\begin{equation}
	P_\sigma^{(q)}(\vecr,\vecr')=\rho^{(q)}_\sigma(\vecr')
-\left|\rho_{\sigma\sigma}^{(q)}(\vecr,\vecr')\right|^2/\rho^{(q)}_\sigma(\vecr) ,
\end{equation}
where, in the mean-field calculations,
\begin{equation}
	\rho_{\sigma\sigma'}(\vecr,\vecr')\equiv
	\sum_{i>0} V_i^*(\vecr\sigma) V_i(\vecr'\sigma') ,
\end{equation}
and $\rho_\sigma(\vecr)=\rho_{\sigma\sigma}(\vecr,\vecr)$.
Again, hereafter in this section,
the superscript $(q)$ is omitted for simplicity.
Let us rewrite $P_\sigma(\vecr,\vecr')=P_\sigma(\vecR,\mathbf{s})$
in terms of the average and the relative positions, $\vecR=(\vecr+\vecr')/2$
and $\mathbf{s}=\vecr-\vecr'$,
then, perform a spherical averaging over the angles of $\mathbf{s}$.
Finally, we expand the $P_\sigma(\vecr,s)$ with respect to $s$ as
\begin{eqnarray}
	P_\sigma(\vecr,s)&\approx& \frac{1}{3} \left(
	\tau_\sigma - \frac{1}{4} \frac{(\mathbf{\nabla}\rho_\sigma)^2}{\rho_\sigma}
	-\frac{\mathbf{j}_\sigma^2}{\rho_\sigma} 
	\right) s^2 \nonumber \\
	&\equiv& \frac{1}{3} D_\sigma(\vecr) s^2
\end{eqnarray}
Then, the localization function $C_\sigma(\vecr)$ is defined as
$C_\sigma(\vecr)=[1+\{ D_\sigma(\vecr)/\tau^{\rm TF}(\vecr)\}^2]^{-1}$,
where the Thomas-Fermi kinetic density
$\tau^{\rm TF}_\sigma(\vecr)=3(6\pi^2)^{2/3}\rho_\sigma^{5/3}$
is introduced to make $C_\sigma(\vecr)$ dimensionless.
$P_\sigma(\vecr,s)\rightarrow 0$ at $s\rightarrow 0$
is guaranteed by the Pauli exclusion principle.
The definition restricts the range of $C_\sigma(\vecr)$ as
$0<C_\sigma(\vecr)\leq 1$.
It is apparent that
the smaller the conditional probability $P_\sigma(\vecr,s)$ is,
the larger the localization function $C_\sigma(\vecr)$.
In other words, $C_\sigma^{(q)}(\vecr)\approx 1$
indicates very little probability of finding two
nearby nucleons with the same spin $\sigma$ and isospin $q$
around the position $\vecr$.
We should emphasize that the localization function
$C_\sigma^{(q)}(\vecr)$ is, in fact, a ``delocalization'' measure of
the same kind of nucleons.
The presence of the $\alpha$ particle requires 
localization of nucleons
with different spins and isospins,
which cannot be quantified by $C_\sigma^{(q)}(\vecr)$.

\section{\label{sec:results} Numerical results}

\subsection{Numerical details}

In the present paper,
instead of full Hartree--Fock--Bogoliubov (HFB) theory, 
we adopt the HF+BCS theory.
It simplifies the numerical computation and
allows us to examine the effect of the pairing
in the local $\alpha$-removal strength.
We truncate the model space for
the pairing correlations.
This is introduced by the number of single-particle orbitals.
For instance, for Sn isotopes,
82 neutron orbitals obtained in the HF+BCS are adopted
for the neutron sector,
while the protons are in the normal phase $(\Delta_p=0$)
with 50 fully occupied orbitals.
The neutron pairing gaps are determined by
the third-order mass difference using the atomic mass evaluation
\cite{AME2021-1,AME2021-2}:
$\Delta_n=1.4$, 1.2, 1.4, and 1.3 MeV for
$A=112$, 116, 120, and 124, respectively.

We use the Skyrme energy density functional with
the SkM$^\ast$ parameter set \cite{Bart82}.
We adopt the three-dimensional (3D) Cartesian grid representation of the square box,
using the computer code developed
in Refs.~\cite{NY05,Eba10,ENI14}.
The 3D grid size is set to be $(1.0\mbox{ fm})^3$.
We adopt all the grid points inside a sphere of the radius of $R=12$ fm.
The differentiation is evaluated with the nine-point finite difference.
The center-of-mass correction is taken into account by
modifying the nucleon's mass as $m\rightarrow m\times A/(A-1)$.
The Coulomb potential is calculated
by solving the Poisson equation with
the conjugate-gradient method,
in which the boundary values are constructed with
the multipole expansion \cite{FKW78}.
The single-particle orbitals are calculated
with the imaginary-time method \cite{DFKW80}.
The iteration is carried out until
the self-consistent solution is obtained.

\subsection{Even-even Sn isotopes}
\begin{figure}[t]
	\centerline{
	\includegraphics[width=0.5\textwidth]{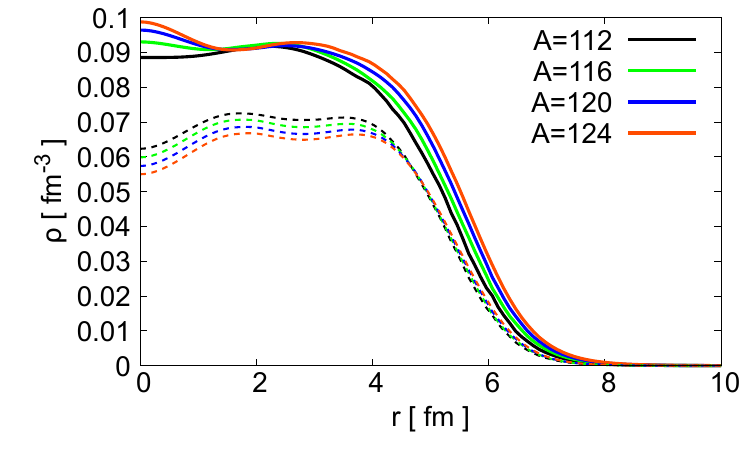}
}
	\caption{
		Nucleon density distributions for neutrons (solid lines)
		and protons (dashed) for Sn isotopes ($A=112$, 116, 120, and 124).
	}
	\label{fig:rho_Sn}
	\medskip
	\centerline{
	\includegraphics[width=0.5\textwidth]{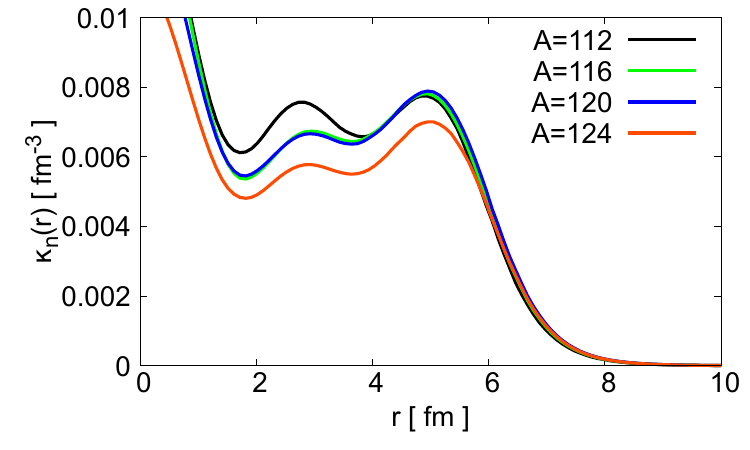}
}
	\caption{
		Neutron pair density distributions 
		for Sn isotopes ($A=112$, 116, 120, and 124).
		The proton pair density vanishes for these isotopes.
	}
	\label{fig:kappa_Sn}
\end{figure}

Since we neglect the rearrangement of the mean fields,
the method is suitable for heavy nuclei
in which the mean-field potentials are relatively stable
against the removal of an $\alpha$ particle (two protons and two neutrons).
Since the ground states of Sn isotopes ($Z=50$) represent
a typical example of pair-rotational bands in spherical nuclei \cite{BB05},
the mean fields should be stable with respect to
the two-neutron removal.
In contrast,
the two-proton removal is expected to have a certain impact
on the mean fields,
because $Z=50$ is a spherical magic number for protons.
Nonetheless,
Cd isotopes ($Z=48$) exhibit typical excitation spectra
of spherical vibrator \cite{BM75}.
Thus, it is meaningful to compare the magnitude of
the local $\alpha$-removal strengths
for different Sn isotopes ($^A$Sn$\rightarrow^{A-4}$Cd).

\subsubsection{Normal density and pair density}

First, let us show the density distributions
for $^{112,116,120,124}$Sn in Fig.~\ref{fig:rho_Sn}.
The neutron radius increases as a function of the neutron number,
while the proton radius stays almost constant.
A dip in the central proton density can be understood
as a shell effect because of the full occupation of
the high-$j$ ($g_{9/2}$) orbital.
As we can expect, the neutron skin develops as increasing the neutron number.
The neutron skin should have an impact on the $\alpha$-particle formation
properties.
Reference~\cite{Typ14}, using the Thomas-Fermi approximation,
gave the $\alpha$-particle density in the surface region
which decreases as the neutron skin increases.
The $\alpha$-cluster formation is also predicted to have
a negative impact on neutron skin thickness.

In Fig.~\ref{fig:kappa_Sn},
the neutron pair densities are shown.
In the present calculation, the central peak at $r\approx 0$
exists,
which may be due to the monopole pairing interaction used in the BCS treatment and may depend on the type of pairing interaction.
The surface peak is located at $r\approx 5$ fm,
whose shape is similar to each other.
Since the proton number $Z=50$ is magic,
the pair density vanishes for protons.

\subsubsection{Local \texorpdfstring{$\alpha$}{alpha}-removal strengths}
\begin{figure}[t]
	\centerline{
	\includegraphics[width=0.45\textwidth]{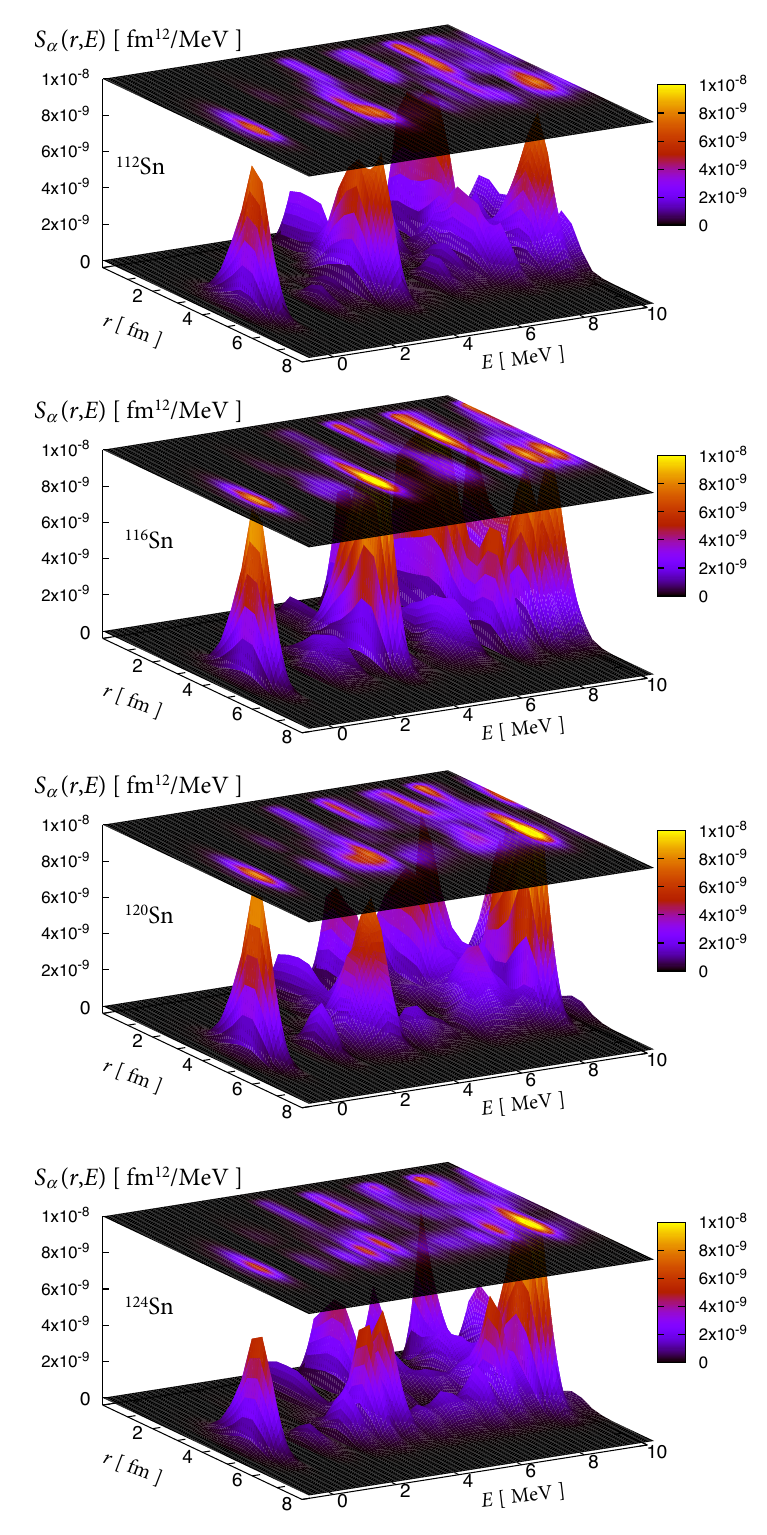}
}
	\caption{
		Local $\alpha$-removal strengths $S_\alpha(r,E)$
		for Sn isotopes ($A=112$, 116, 120, and 124).
		The energy $E$ corresponds to the excitation
		energy of the residual nuclei after the removal
		of an $\alpha$ particle at the radial position $r$.
		The discrete strengths are smeared by
		Gaussians with a width of 100 keV.
	}
	\label{fig:SrE_Sn}
\end{figure}

Since the numerical calculation is performed with
the vanishing boundary condition, all the quasiparticle energies
are discrete.
To visualize the local $\alpha$-removal strength
$S_\alpha(r,E)$ as a function of excitation energy $E$,
we replace the $\delta$ function in Eq. (\ref{eq:S(r,E)_MF}) by
the Gaussian function of the width of $\gamma=100$ keV.
The calculated local $\alpha$-removal strengths for Sn isotopes
are shown in Fig.~\ref{fig:SrE_Sn}.
For each isotope,
there is an isolated peak corresponding to the ground-ground transition
($E=0$).
This $\alpha$-removal strength to the ground state is located near the surface region.
At excitation energies of $E \gtrsim 3$ MeV,
there are peaks whose magnitude is comparable to or even larger than
the transitions to the ground state.
In contrast to the ground-ground transition,
the strengths are not only in the surface region,
but also in the interior region with $r<3$ fm.
This indicates that the $\alpha$ particle may exist
deep inside the nucleus.
However, in the $\alpha$-knockout reaction, these $\alpha$ particles
are difficult to come out of the nucleus because of the strong absorption.
No strength is shown at $r=0$ in Fig.~\ref{fig:SrE_Sn}
that shows $S_\alpha(r,E)$ in the range of $E<10$ MeV.
This is because the proton amplitude vanishes at $r=0$, $F_k^{(p)}(0)=0$.
The binding energy of the proton $s_{1/2}$ state is
larger than the $g_{9/2}$ state, by more than 20 MeV.
Thus, nonzero proton amplitude at the center $F_k^{(p)}(0)\neq 0$
appears only for $E > 40$ MeV.

\subsubsection{Residual nuclei in the ground state}

\begin{figure}[t]
	\centerline{
	\includegraphics[width=0.5\textwidth]{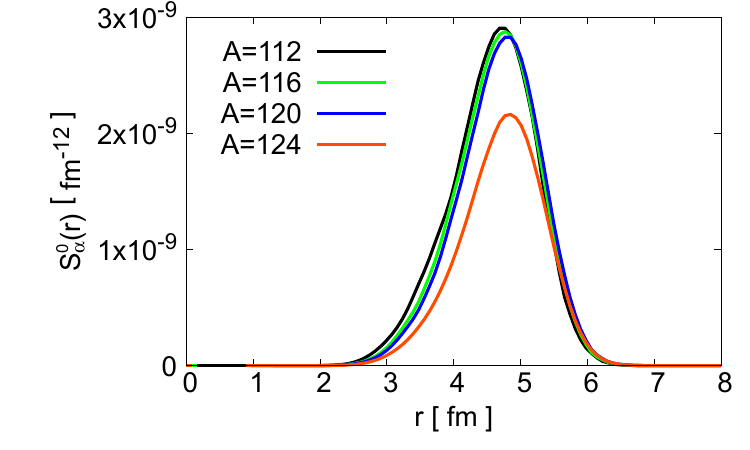}
}
	\caption{
		Local $\alpha$-removal strength to the ground state ${\cal S}_\alpha^0(\vecr)$
		for Sn isotopes ($A=112$, 116, 120, and 124).
	}
	\label{fig:S0_Sn}
\end{figure}

To examine the structure of the local $\alpha$-removal strengths
to the ground state of the residual nuclei,
in Fig.~\ref{fig:S0_Sn},
we show the strength of Eq. (\ref{eq:S_EDeltaE})
with $E=0$ and $\Delta E\rightarrow 0+$,
\begin{eqnarray}
	{\cal S}_\alpha^0 (\vecr) &\equiv&
	{\cal S}_\alpha(\vecr)_{E=0,\Delta E=2\epsilon}
	\nonumber \\
	&=&\int_{-\epsilon}^\epsilon S_\alpha(\vecr,E) dE
	=F_0^{(n)}(\vecr) F_0^{(p)}(\vecr) ,
	\label{eq:S^0}
\end{eqnarray}
where $\epsilon$ is a positive infinitesimal.
When we remove an $\alpha$ particle at the position $\vecr$,
${\cal S}_\alpha^0(\vecr)$ can be regarded as a quantity proportional to
the probability that the residual nucleus becomes the ground state.
The shape of the peak is almost identical among these isotopes
and located at $r=|\vecr|\approx 4.7$--$4.8$ fm.
This position approximately corresponds to the position $r$ that gives $\rho(\vecr)=(2/3)\times\rho(\bm{0})$
(Fig.~\ref{fig:rho_Sn}).
It is near the surface, however,
the radial value $r$ is significantly smaller than
the peak position of the $\alpha$ density $n_\alpha(\vecr)$
predicted in Ref.~\cite{Typ14}.
In fact, the peak position
of the $\alpha$ density $n_\alpha(\vecr)$ in Ref.~\cite{Typ14}
is located at $6.5 < r < 7.5$ fm,
which roughly corresponds to the value $r$ with 
$\rho_q(\vecr)\approx \rho_q(\bm{0})/10$.
The $\alpha$ density $n_\alpha(\vecr)$ is predicted to
vanish in the region of $r<6$ fm
for Sn isotopes \cite{Typ14}.

The peak height is similar to each other for $^{112,116,120}$Sn,
while it is apparently smaller for $^{124}$Sn.
This is naturally understood by the pair density in Fig.~\ref{fig:kappa_Sn}.
The proton matrix elements
$F_0^{(p)}(\vecr)$, which is given by Eq.~(\ref{eq:F_0_normal}),
are determined by the HOO, namely $g_{9/2}$ orbitals.
They are surface peaked and
approximately identical to each other among all the isotopes.
The neutron matrix element $F_0^{(n)}(\vecr)$ is given by
$|\kappa_n(\vecr)|^2$,
according to Eq.~(\ref{eq:F_0_super}).
Therefore, variations in
${\cal S}_\alpha^0(\vecr)=F_0^{(n)}(\vecr)F_0^{(p)}(\vecr)$
come from those in $F_0^{(n)}(\vecr)=|\kappa_n(\vecr)|^2$.
A reduction in $\kappa_n(\vecr)$ at $r\approx 5$ fm
is the reason of the reduced peak height in ${\cal S}_\alpha^0(\vecr)$ in $^{124}$Sn.
This is easily confirmed by artificially increasing the neutron pairing gap:
We have found that the peak height of ${\cal S}_\alpha^0(\vecr)$
increases by about 50 \%
when we double the pairing gap $\Delta_n$.

The $\alpha$-knockout experimental data in Ref.~\cite{Tan21}
clearly indicate a monotonic decrease as a function of the neutron number.
The experiment measures the missing-mass spectra to extract
the cross section in which the residual nucleus is in
the ground state.
Therefore, this isotopic dependence should be related to ${\cal S}_\alpha^0(\vecr)$
at the nuclear surface.
The peak height of ${\cal S}_\alpha^0(\vecr)$ shown in Fig.~\ref{fig:S0_Sn}
is similar to each other except for $^{124}$Sn.
Furthermore, they are almost identical at $r\gtrsim 5.5$ fm
where the $\alpha$ knockout mainly takes place, namely,
\begin{equation}
	{\cal S}_\alpha^0(\vecr)_{A=112}\approx
	{\cal S}_\alpha^0(\vecr)_{A=116}\approx
	{\cal S}_\alpha^0(\vecr)_{A=120}\approx
	{\cal S}_\alpha^0(\vecr)_{A=124} ,
\end{equation}
at $r\gtrsim 5.5$ fm.
In other words, ${\cal S}_\alpha^0(\vecr)$ is universal for these isotopes
in the surface region.
This seems to be inconsistent with the experimental observation,
at first sight.

However, we need to further examine the relationship between
the cross section and the local $\alpha$-removal strength ${\cal S}_\alpha^0(\vecr)$.
Since there is a strong absorption of the $\alpha$ particle
inside the nucleus,
the cross section may not be correlated with the values at the same $r$,
but we should compare those at a fixed value of nucleon density for each isotope.
The nuclear radii apparently increase as the neutron number increases,
because of the neutron skin effect (Fig.~\ref{fig:rho_Sn}),
namely $R_{112}<R_{116}<R_{120}<R_{124}$.
Thus, the ${\cal S}_\alpha^0(\vecr)$ values at the surface (fixed density)
decrease as a function of the neutron number:
\begin{equation}
	{\cal S}_\alpha^0(\vecR_{112})>
	{\cal S}_\alpha^0(\vecR_{116})>
	{\cal S}_\alpha^0(\vecR_{120})>
	{\cal S}_\alpha^0(\vecR_{124}) .
	\label{eq:S^0_ordering}
\end{equation}
Therefore, the universal behavior of ${\cal S}_\alpha^0(\vecr)$ may be
consistent with the experimental observation.

\begin{figure}[t]
	\centerline{
	\includegraphics[width=0.5\textwidth]{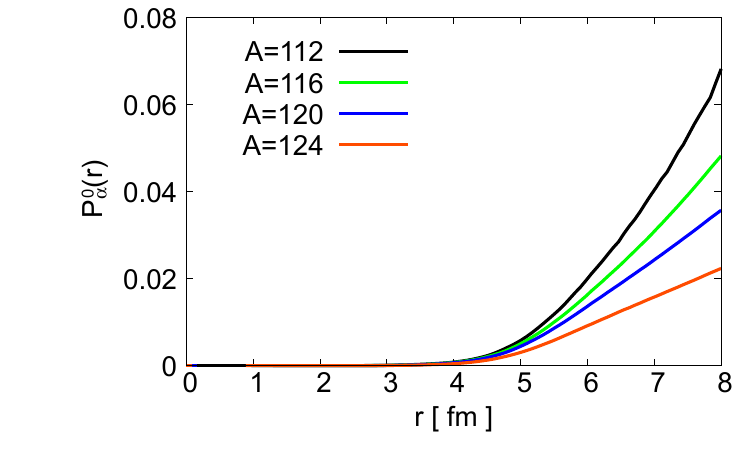}
}
	\caption{
		Local $\alpha$ probability $P_\alpha^0(\vecr)$
		for Sn isotopes ($A=112$, 116, 120, and 124).
	}
	\label{fig:P0_Sn}
\end{figure}

To visualize this neutron number dependence,
we define a dimensionless quantity,
``local $\alpha$ probability,'' as the ${\cal S}_\alpha^0(\vecr)$ value relative
to the density.
\begin{equation}
	P_\alpha^0(\vecr)\equiv \frac{{\cal S}_\alpha^0(\vecr)}
	{\rho_{n\uparrow}(\vecr)
	\rho_{n\downarrow}(\vecr)
	\rho_{p\uparrow}(\vecr)
	\rho_{p\downarrow}(\vecr)} .
	\label{eq:P^0}
\end{equation}
$P_\alpha^0(\vecr)$ can be regarded as
the probability to find an $\alpha$ particle at the position $\vecr$
under the condition that the residual nucleus is in the ground state,
normalized to the probability of finding the four kinds of nucleons.
The local $\alpha$ probability is plotted in Fig.~\ref{fig:P0_Sn}.
$P_\alpha^0(\vecr)$ clearly indicates the monotonic
decrease as the neutron number,
which is consistent with the experiment \cite{Tan21}.

\subsubsection{Excited residual nuclei}

\begin{figure}[t]
	\centerline{
	\includegraphics[width=0.5\textwidth]{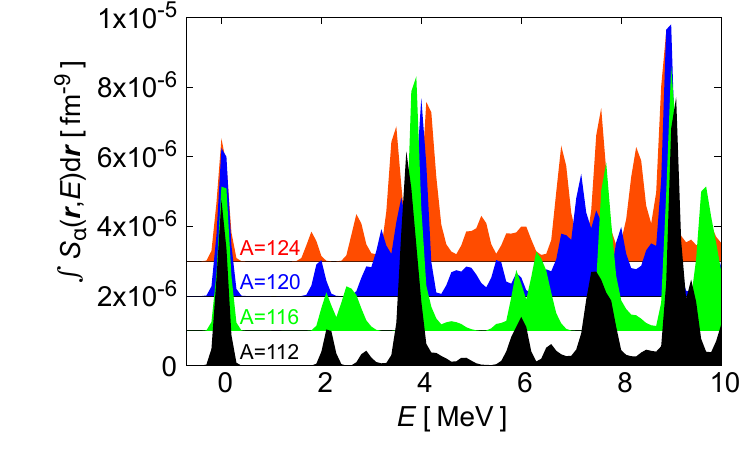}
}
	\caption{
		Integrated local $\alpha$-removal strength
		$\displaystyle\int S_\alpha(\vecr,E) d\vecr$,
		for Sn isotopes ($A=112$, 116, 120, and 124).
		Those for $A=116$, 120, and 124 are shifted upwards
		by $10^{-6}$, $2\times10^{-6}$, and $3\times10^{-6}$ fm$^{-9}$,
		respectively.
	}
	\label{fig:E-S_Sn}
\end{figure}

The local $\alpha$-removal strength, in principle, contains
information on $\alpha$ knockout to excited residual nuclei.
Since the excited states are simply given by neutron 2qp states
and proton particle-hole excitations,
we should keep in mind that it is a qualitative measure.
In Fig.~\ref{fig:E-S_Sn}, $S_\alpha(\vecr,E)$ integrated over
the space $\vecr$ is shown for Sn isotopes.
The small peak next to the ground state ($E\approx 2$ MeV)
corresponds to proton excitation in which
one of the protons is removed from the $g_{9/2}$ orbit
and the other from $p_{1/2}$.
The $\alpha$-removal strengths to some excited states of residual nuclei are as strong as those to the ground state.

\begin{figure}[t]
	\centerline{
	\includegraphics[width=0.5\textwidth]{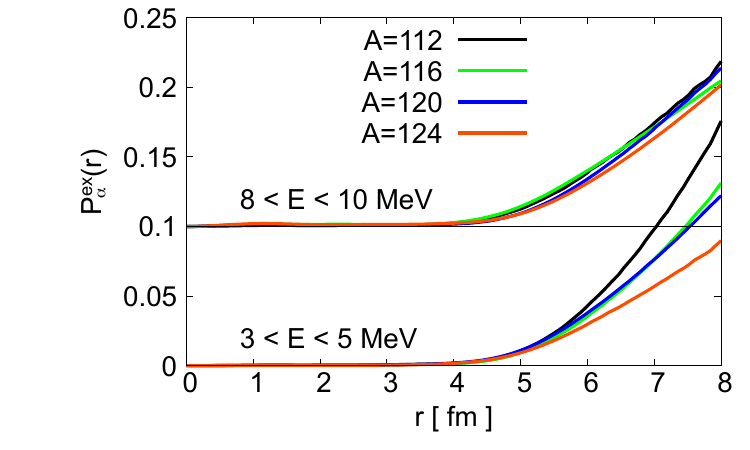}
}
	\caption{
		Local $\alpha$ probabilities for excited residual nuclei
		in Sn isotopes ($A=112$, 116, 120, and 124),
		which are defined as Eq.~(\ref{eq:P^ex}) with
		$E=4$ MeV and $\Delta E=2$ MeV.
		Those with $E=9$ MeV
		are shifted upwards by $0.1$.
	}
	\label{fig:Pex_Sn}
\end{figure}

It may be of interest to investigate the structure of
the local $\alpha$ probability when the residual nuclei are excited.
Since there are two prominent peaks in Fig.~\ref{fig:E-S_Sn},
one around $E\approx 4$ MeV and the other around 9 MeV,
we set $E=4$ (9) MeV and $\Delta E=2$ MeV,
to calculate the local $\alpha$ probability as
\begin{equation}
P_\alpha^{\rm ex}(\vecr)_{E,\Delta E}\equiv
	\frac{{\cal S}_\alpha(\vecr)_{E,\Delta E}}
	{\rho_{n\uparrow}(\vecr)
	\rho_{n\downarrow}(\vecr)
	\rho_{p\uparrow}(\vecr)
	\rho_{p\downarrow}(\vecr)} ,
\label{eq:P^ex}
\end{equation}
where
${\cal S}_\alpha(\vecr)_{E,\Delta E}$
is given by Eq. (\ref{eq:S_EDeltaE}).
These are shown in Fig.~\ref{fig:Pex_Sn}.
The local $\alpha$ probabilities for excited residual nuclei
are enhanced in the low-density region.
The monotonic increase as a function of $r$ is the same as
those to the ground state $P_\alpha^0(\vecr)$,
and seems to be universal.
However, their isotopic dependence is not as prominent as $P_\alpha^0(\vecr)$.
$P_\alpha^{\rm ex}(\vecr)_{E,\Delta E}$ with $E=9$ MeV and $\Delta E=2$ MeV
for different isotopes
are similar to each other.
Since we neglect the effects of the rearrangement and the collective states,
these numbers should not be taken seriously.
Nevertheless, this may suggest that the $\alpha$-knockout reaction
with excited residual nuclei may not show
the prominent neutron number dependence,
in contrast with those for the ground residual nuclei.

\begin{figure}[t]
	\centerline{
	\includegraphics[width=0.5\textwidth]{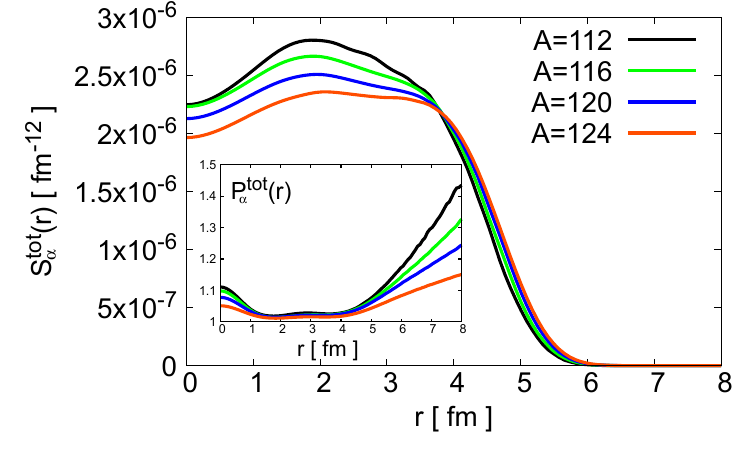}
}
	\caption{
		Energy-integrated local $\alpha$-removal strength
		for Sn isotopes ($A=112$, 116, 120, and 124).
	}
	\label{fig:Stot_Sn}
\end{figure}

Integrating over the entire energy range,
the total strength of the local $\alpha$-removal can be easily estimated
in the mean-field approximation as
\begin{equation}
	S_\alpha^{\rm tot}(\vecr)=
	\int_{-\infty}^\infty S_\alpha(\vecr,E) dE
	=
	S_{\uparrow\downarrow}^{(n)}(\vecr)
	S_{\uparrow\downarrow}^{(p)}(\vecr) ,
	\label{eq:S_tot}
\end{equation}
where 
\begin{eqnarray}
	S_{\uparrow\downarrow}^{(q)}(\vecr) &=& 
	\bra{\Phi_0^{N_q}}
	\psi^\dagger_{q\downarrow}(\vecr)
	\psi^\dagger_{q\uparrow}(\vecr)
	\psi_{q\uparrow}(\vecr)
	\psi_{q\downarrow}(\vecr)
	\ket{\Phi_0^{N_q}} \nonumber \\
	&=&\rho^{(q)}_{\uparrow}(\vecr)
	\rho^{(q)}_{\downarrow}(\vecr)
	-\left| \rho^{(q)}_{\uparrow\downarrow}(\vecr) \right|^2
	+\left| \kappa^{(q)}(\vecr)\right|^2 .
	\label{eq:S_updown}
\end{eqnarray}
The quantity of Eq. (\ref{eq:S_tot}) can be written as
$
S_\alpha^{\rm tot}(\vecr)=\bra{\Phi_0^A}
\alpha^\dagger(\vecr)\alpha(\vecr) \ket{\Phi_0^A}
$,
which may be regarded as the $\alpha$-particle density
distribution.
This is shown in Fig.~\ref{fig:Stot_Sn}.
The major contribution to $S_\alpha^{\rm tot}(\vecr)$
is the first term of Eq. (\ref{eq:S_updown}),
which is a local density product of nucleons with spin up and down.
Thus, $S_\alpha^{\rm tot}(\vecr)$ of Eq.~(\ref{eq:S_tot})
mainly comes from a trivial density product of four kinds of nucleons.
This is nothing but the denominator of
Eqs. (\ref{eq:P^0}) and (\ref{eq:P^ex}).
If we normalize $S_\alpha^{\rm tot}(\vecr)$ with respect to
this trivial density product factor,
\begin{equation}
P_\alpha^{\rm tot}(\vecr)\equiv
	\frac{S_\alpha^{\rm tot}(\vecr)}
	{\rho_{n\uparrow}(\vecr)
	\rho_{n\downarrow}(\vecr)
	\rho_{p\uparrow}(\vecr)
	\rho_{p\downarrow}(\vecr)} ,
\label{eq:P_tot}
\end{equation}
we obtain results shown in the inset of Fig.~\ref{fig:Stot_Sn}.
Again, in the surface region, we observe the clear neutron-number dependence
the same as $P_\alpha^0(\vecr)$ in Fig.~\ref{fig:P0_Sn}.

If we neglect the second and the third terms in Eq.~(\ref{eq:S_updown}),
we trivially have $P_\alpha^{\rm tot}(\vecr)=1$.
Since the second term of Eq. (\ref{eq:S_updown})
vanishes for the time-even ground state,
the enhancement effect is due to the third term,
namely, the effect of neutron pairing.
Therefore, the surface $\alpha$ formation may be
understood as the fact that the pair density distribution
is more extended than the normal density.

\subsubsection{Localization function}

\begin{figure}
	\centerline{
	\includegraphics[width=0.5\textwidth]{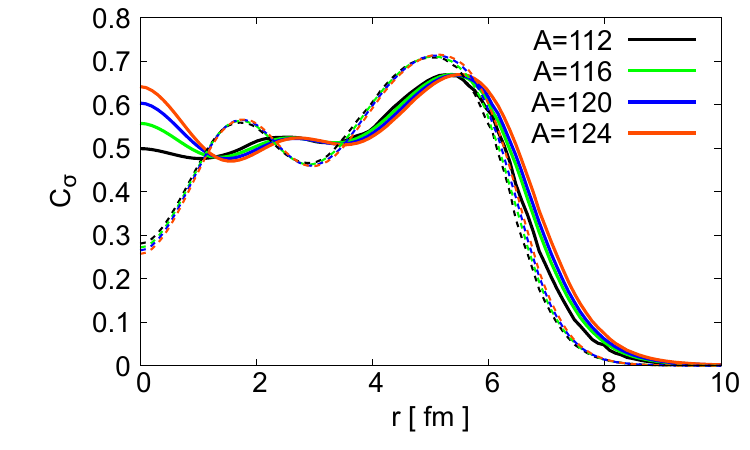}
}
	\caption{
		Localization functions for neutrons (solid lines)
		and protons (dashed lines) for Sn isotopes ($A=112$, 116, 120, and 124).
		The values with spin up ($\sigma=+1/2$) $C_{+1/2}$ are shown
		in the figure,
		but those for spin down ($\sigma=-1/2$) are identical.
	}
	\label{fig:Cs_Sn}
\end{figure}

Before closing this section,
we examine the validity of the localization function.
The calculated localization function  $C_\sigma(\vecr)$
for Sn isotopes is presented in Fig.~\ref{fig:Cs_Sn}.
$C_\sigma(\vecr)$ are approximately identical for all the isotopes. 
We observe a bump in $C_\sigma(\vecr)$ at $r\approx 5$ fm for protons and at
$r\approx 5.5$ fm for neutrons.
These values $r$ of the peak positions are
larger than those of ${\cal S}_\alpha^0(\vecr)$
(Fig.~\ref{fig:S0_Sn}).
Besides, the profile of the function is significantly different
between $C_\sigma(\vecr)$ and ${\cal S}_\alpha^0(\vecr)$.
In comparison with the summed local $\alpha$-removal strength $S_\alpha^{\rm tot}(\vecr)$
in Fig.~\ref{fig:Stot_Sn},
we again observe significantly different peak positions and profiles.
There is no surface peak, and
the peak structure almost disappears in Fig.~\ref{fig:Stot_Sn}.
Therefore, it could be misleading to identify
the localization function $C_\sigma(\vecr)$
as the indicator of the $\alpha$-particle formation
in the mean-field theory.

\section{Conclusion}
\label{sec:conclusion}

To quantify the $\alpha$-particle formation,
the local $\alpha$-removal strength $S_\alpha(\vecr,E)$ is proposed.
When we remove an $\alpha$ particle at the position $\vecr$ from a nucleus,
the final state in the residual nucleus can be expanded
in the energy eigenstates.
The local $\alpha$-removal strength $S_\alpha(\vecr,E)$
corresponds to the strength
to produce the state at an energy $E$ in the residual nucleus.
This quantity is defined with respect to a many-body wave function,
thus, it can be calculated using various quantum many-body techniques,
in principle.
The calculation becomes manageable
when we adopt some approximations,
such as the mean-field approximation
(energy density-functional theory).
Furthermore, if we neglect the rearrangement of the mean fields
after the removal of an $\alpha$ particle,
the computational cost necessary for the local $\alpha$-removal strength
is less than that for the mean-field calculation to obtain the ground state.
We use these approximations in the present paper.

We calculate the local $\alpha$-removal strengths for Sn isotopes
with $A=112$, 116, 120, and 124.
These nuclei are studied by a recent $\alpha$-knockout experiment,
in which the cross sections with the residual nuclei in the ground state
clearly indicate a monotonically decreasing
trend as a function of the neutron number.
This prominent neutron-number dependence is not found in the $\alpha$-removal strength to
the ground state, ${\cal S}_\alpha^0(\vecr)$, of Eq. (\ref{eq:S^0}).
In fact, the function ${\cal S}_\alpha^0(\vecr)$ is almost universal in the surface
for all these isotopes.
Nevertheless, the observed neutron-number dependence
is well reproduced by the local $\alpha$ probability, $P_\alpha^0(\vecr)$,
of Eq. (\ref{eq:P^0}).
The monotonic decrease as a function of the neutron number
is especially evident at the nuclear surface of $r\gtrsim 6$ fm,
where the $\alpha$-knockout reaction is supposed to take place.

It is also possible to explain the experimental trend using the
universal character of ${\cal S}_\alpha^0(\vecr)$ in the surface region
together with the development of the neutron skin. Since the absorption of
the $\alpha$ particle is strong, the knockout reaction is allowed only
in the low-density region. This means that the radial value $r$ of the
region probed by the $\alpha$-knockout reaction is an increasing
function of the neutron number. Therefore, the ${\cal S}_\alpha^0(\vecr)$
values relevant to the knockout cross section decrease with the
neutron number. This naturally explains the experimental data.

To identify the $\alpha$ cluster,
Ref.~\cite{KHME22} proposed criteria combining the localization 
function with the compactness of the density localization.
It may be useful for light deformed nuclei which show a
prominent density localization.
Apparently, the Sn isotopes in the present study
satisfy none of these criteria;
the ground states are spherical and show no density localization.
On the other hand, the calculated local $\alpha$-removal strength
to the ground state shows a surface peak structure
and indicates the importance of the pair density.
Since the pair density dominates
over the normal density in the nuclear surface region,
the last term of Eq.~(\ref{eq:S_updown})
should play a crucial role in the surface $\alpha$ formation.

The local $\alpha$-removal strength in the present approximations
can be calculated with a single state.
In other words, the states in the residual nucleus are not constructed
explicitly.
Therefore,
with a proper choice of the mean-field Hamiltonian,
it can be evaluated in a time-dependent manner
with the time-dependent density-functional theory (TDDFT).
Recently, the nuclear TDDFT calculations have been renovated
to include the pair density \cite{ASC08,Eba10,SBMR11,Has12,ENI14,HS16,MSW17}.
It is of significant interest to investigate
the $\alpha$-particle formation probability during nuclear reactions,
such as heavy-ion reactions, fusion, and fission.

In the present paper, we introduce several approximations for
feasibility of the numerical computation.
The BCS approximation can be lifted and replaced by the full HFB calculation.
It may be of interest to study how the type of the pairing interaction,
such as volume or surface or mixed types,
affects the $\alpha$-formation properties.
We also neglect the rearrangement of the
mean fields before and after the removal of an $\alpha$ particle.
The numerical calculation is extremely feasible with this approximation.
However, it is a drastic approximation even for heavy nuclei.
Especially, near the doubly closed nuclei,
the nuclear shape may be changed, and the approximation may not be justified.
To improve this,
the calculation with proper treatment of the rearrangement
is currently under progress.
This may lead to a quantitative evaluation of
the $\alpha$-knockout cross section.
Furthermore,
the inclusion of the proton-neutron pairing is an interesting
subject in the future.

\begin{acknowledgments}
This work is supported in part by JSPS KAKENHI Grants No. JP18H01209,
No. JP19H05142, No. JP23H01167, No. JP20K03964, and No. JP19KK0343.
This research in part used computational resources provided
by Multidisciplinary Cooperative Research Program in the Center for
Computational Sciences, University of Tsukuba.
\end{acknowledgments}

\bibliographystyle{apsrev4-2}
\bibliography{local_alpha}

\end{document}